# Raman spectroscopy study of rotated double-layer graphene: Misorientation-angle dependence of electronic structure

Kwanpyo Kim[1,2,3], Sinisa Coh[1,3], Liang Z. Tan[1,3], William Regan[1,3], Jong Min Yuk[1,3,4], Eric Chatterjee[1], M. F. Crommie[1,2,3], Marvin L. Cohen[1,2,3], Steven G. Louie[1,3], and A. Zettl[1,2,3]

*[1]Department of Physics and [2]Center of Integrated Nanomechanical Systems, University of California at Berkeley, [3]Materials Sciences Division, Lawrence Berkeley National Laboratory, Berkeley, CA 94720, U.S.A. and [4]Department of Materials Science and Engineering, KAIST, Daejeon 305-701, Korea.*

**We present a systematic Raman study of unconventionally-stacked double-layer graphene, and find that the spectrum strongly depends on the relative rotation angle between layers. Rotation-dependent trends in the position, width and intensity of graphene 2D and G peaks are experimentally established and accounted for theoretically. Our theoretical analysis reveals that changes in electronic band structure due to the interlayer interaction, such as rotational-angle dependent Van Hove singularities, are responsible for the observed spectral features. Our combined experimental and theoretical study provides a deeper understanding of the electronic band structure of rotated double-layer graphene, and leads to a practical way to identify and analyze rotation angles of misoriented double-layer graphene.**

Recently there has been growing interest in double-layer graphene in which the two graphene layers are not conventionally stacked but relatively rotated by an arbitrary angle [1-12]. Such graphene double layers are expected to display characteristics distinct from both monolayer graphene as well as the extensively studied AB-stacked bilayer graphene [13-16]. Previous theoretical investigations suggest that electronic and optical properties of double layer graphene will strongly depend on this rotational angle [1-5]. Because the entire range of rotational angles is in principle experimentally accessible via artificial stacking, the properties of rotated double-layer graphene might be tuned to suit the application at hand, making this material a useful component in future nano-electronic devices. Limited low-energy electrical transport measurements have suggested that rotated graphene layers maintain the linear dispersion relation as in single-layer graphene [6,7]. Furthermore, angle-resolved photoemission spectroscopy measurement has shown that rotated layers in multilayer epitaxial graphene exhibit weak interlayer interactions [8]. On the other hand, scanning tunnelling microscopy studies in the low rotation angle regime have demonstrated strong interlayer interactions such as carrier velocity renormalization and the occurrence of Van Hove singularities away from the Dirac point energy [9,10]. Despite these suggestive findings and the applications potential, there have unfortunately been no comprehensive experimental studies of the influence of rotation angle on the electronic properties of double-layer graphene.

In this Letter we present a systematic experimental and theoretical study of rotated double-layer graphene. We employ Raman spectroscopy, a powerful tool for investigating the electronic and vibrational properties of carbon-based materials [17-20], together with theoretical calculations of the electronic-structure-dependent Raman response. We experimentally sample a range of misorientation angles from 0 to 30 degrees in steps of ~1 degree, and we focus on the intensity, peak position, and peak width of the 2D and G Raman modes. Previous limited Raman studies on folded graphene [21-25] have made interesting observations relevant to rotated double-layer graphene, such as 2D peak blue-shifts and G peak resonance, but the origin of these phenomena could not be clearly identified.

To obtain rotated double-layer graphene we start by synthesizing monolayer polycrystalline graphene via chemical vapor deposition (CVD), yielding material with a grain size of several micrometers [26-28]. By consecutively transferring two such monolayers of CVD graphene onto a transmission electron microscope (TEM) grid [29], we obtain double-layer graphene with domains having randomly rotated stacking angles. We utilize holey carbon TEM grids with an array of holes 2 μm in diameter. Within each hole we perform a TEM analysis to identify the misorientation angle of the suspended double-layer.

Fig. 1a shows a typical real-space TEM image of double-layer graphene. The sample areas are largely clean and flat. There are also apparent small scattered dark regions, which are not present in single-layer graphene [28]. These dark regions are most likely carbon residues trapped between the two graphene layers and can be used to quickly distinguish double-layer from single-layer graphene regions in our samples. Other than these small residue pockets, the rotated double-layer graphene samples generally have clean interfaces between the two layers, showing clear moiré patterns at the atomic scale (Fig. 1c).

For this study, we choose only double-layer graphene specimens where each layer suspended over the hole consists of a single domain. For these samples, two sets of hexagonal diffraction spots, one from each layer, are obtained, as exemplified in Fig. 1b. From the diffraction spots we unambiguously determine the rotation angle between the two misoriented single-crystal layers spanning a hole. We are cautious to avoid samples with tilt grain boundaries [27] or local fold structure from the transfer process [30], which would give more than two sets of hexagonal diffraction spots over the sample area. In contrast to previous Raman studies with folded graphene samples, here we can easily mass-produce rotated double-layer samples with broad coverage of rotation angles [28]. Following TEM analysis, we perform Raman spectroscopy measurements on the indexed rotated double-layers inside the designated holes.

Figure 1d shows selected Raman spectra of a misoriented double-layer graphene (having rotation angles of 3, 7, 10, 14, 20, and 27 degrees) and of single-layer graphene with 633 nm laser wavelength (1.96 eV). The single-layer graphene shows the typical signature of a 2D/G peak integral intensity ratio around 6, and 2D peak width (FWHM) of $28.7 \pm 0.9$ cm$^{-1}$. In the case of rotated double-layer graphene, the data clearly show a change in spectral features from low-angle to high-angle misorientations. Low-angle (<~ 8 degrees, for 633 nm laser wavelength) misoriented double-layers exhibit the Raman signature of a strong coupling between layers. In the high-angle regime (> 13 degrees), the double-layers display Raman spectra closer to those of single-layer graphene. The 2D peak of double-layer graphene is blue-shifted relative to the 2D peak in monolayer graphene, with the blue-shift magnitude depending non-monotonically on the rotation angle. We also observe a strong resonance of the G peak at an intermediate angle (~ 10 degrees). Additionally we find that the peak position and width of G peak are almost angle independent, while the intensity and width of the 2D peak again show quite complex angle dependence. As we will show later, these observations can be explained by the rotation-angle dependent electronic bandstructure.

We also observe a D peak (two times weaker than G peak), which indicates the presence of defects, most likely introduced by the fabrication process and TEM characterization. Additionally, at certain rotational angles we observe extra peaks around the G peak. These peaks have been investigated in recent studies [25,31].

We consider here in more detail our experimental results for the G Raman peaks of rotated double-layer graphene. Measured Raman spectra with frequencies close to the G peak are shown in Fig. 2a for a single layer and selected rotational angles of double-layer graphene. Compared to single-layer, the double-layer graphene G peak has slightly larger FWHM (2 to 6 cm$^{-1}$) while the center location shows slight shift (1 to 3 cm$^{-1}$) towards red [28]. Even though these broadening and red-shifts are more prominent at the very-low angle regimes (< 3 degrees), the rotation-angle dependence is generally very weak [28].   The positions of the G peak are consistent and show only slight variations throughout the double-layer samples.   This demonstrates that double-layer samples do not have significant doping or strain differences from single-layer graphene [17] and the effect of trapped carbon residues is minor.   Unlike its width and peak position, the G mode intensity at laser wavelength of 633 nm is, however, extremely angle dependent.   As shown in Fig. 2b we find more than a thirty-fold increase of G peak intensity at 10 and 11 degrees, compared to other angles, where the intensity is largely angle independent.   We also show in Fig. 2c that with larger laser excitation energy of 2.41 eV (514 nm) G peak enhancement occurs at higher angles (~13 degrees). The inset of Fig. 2c shows a linear relation between the experimental laser energy and the rotational angle at which G peak enhancement is observed.

The graphene 2D Raman peak, which is the most sensitive peak to electronic and phonon band structure changes in graphene [17], exhibits even more complex rotational-angle dependence.   Fig. 3a shows the measured Raman spectra around the graphene 2D peak for double-layer and single-layer graphene.   The full width at half maximum (FWHM) of the 2D peak is large at small rotation angles and is close to the single-layer values at large angles (Fig. 3b).   However, the decrease is not monotonic, and around 7 - 9 degrees we observe an increase in FWHM.   The 2D peak position overall shows a blue-shift and is strongly angle dependent (Fig. 3c).   We find from 0 to 8 degrees a decrease in blue-shift, from 10 to 1 cm$^{-1}$, followed by a sharp increase to 22 cm$^{-1}$ at 10 degrees.   From 10 to 17 degrees, the blue-shift decreases to 11 cm$^{-1}$ and is nearly angle independent at higher angles.   We observe that the rotation angle at which there is strongest variation in the 2D peak position coincides with that of the G peak intensity enhancement angle and local increase in 2D FWHM.   This implies that these features share a common origin.

The measured 2D intensity also shows a dramatic change with respect to rotation angles (black square data in Fig. 3d).    For non-interacting double-layer graphene, one would expect twofold increase in the 2D peak intensity as compared to the single-layer graphene.    In the low angle regime (< 10 degrees), however, the integral intensity of the 2D peak is nearly 50 % reduced compared to what would be expected for the non-interacting double-layer graphene.    In the middle range, the intensity shows the increasing trend and finally above 17 degrees become similar to or greater than two times the value in single layer graphene.

In order to achieve a more complete understanding of rotation angle dependence of double-layer graphene Raman features, we turn to a theoretical analysis of the Raman spectrum as a function of rotational angle [28].    We first consider a simplified band structure analysis that accounts for the existence of the experimentally observed critical misorientation angle; we then perform the tight binding calculation of Raman spectrum in the entire range of misorientation angles.

In momentum space, for two noninteracting layers misoriented by angle $\theta$, the respective momentum space Brillouin zones (BZ) of the top and bottom graphene layers are rotated by $\theta$ as shown in Fig. 4a.    Band structure modifications occur mostly where the Dirac cones of top and bottom layers are overlapping.    Fig. 4b shows the simplified electronic band-structure in the vicinity of the overlap of these two Dirac cones.    In this region, the density of states (DOS) of double layer graphene is modified from non-interacting graphene, exhibiting Van Hove singularities as shown in Fig. 4c [9]. The energy difference between conduction and valence Van Hove singularities scales with the rotational angle and can reach the energy of a few eV at the optical range in the case of higher rotational angles.

For a given Raman excitation laser energy $E_{laser}$, we calculate a critical rotational angle $\theta_c$ where the energy between the conduction and valence Van Hove singularities equals $E_{laser}$.    Using the Dirac dispersion relation of monolayer graphene, the critical angle can be calculated as

$$\theta_c = \Delta k / K = \frac{3aE_{laser}}{\hbar v_f 4\pi}, \tag{1}$$

where $a$ is the lattice parameter of graphene (2.46 Å), $\hbar$ is the reduced Planck's constant, and $v_f$ is the Fermi velocity in monolayer graphene ($10^6$ m/s). In the case of a 633 nm laser excitation (1.96 eV), we find $\theta_c = 10°$, in excellent agreement with $\theta_c = 10°$ determined experimentally (Fig. 2b). Therefore, we expect that for double-layer graphene with rotational angle close to $\theta_c$, the Raman spectra at corresponding laser energy $E_{laser}$ will be strongly affected by the coupling between top and bottom graphene layer. The prediction of our simple model also agrees well with the laser-energy dependence of rotational angles at which we observe the G peak enhancement (inset of Fig. 2c). Furthermore, we expect that for angles larger than $\theta_c$ Raman spectra will resemble those of a single layer graphene since all the optical excitation occurs in an isolated simple Dirac cone structure (intervalley 2D scattering process for such a case is shown with blue lines in Fig. 4e). On the other hand, for angles smaller than $\theta_c$ we expect Raman spectra quite different from those of single layer graphene, since the closeness of the Dirac cones from the top and bottom layers in momentum space allows for scattering paths significantly different from those of the monolayer (black lines in Fig. 4e).

Although the simple model utilized above correctly accounts for the observed critical rotation angle, it cannot provide a satisfactory quantitative explanation of the detailed Raman peak intensities observed experimentally (especially below critical mismatch angles) and a more complete theoretical approach is needed. We therefore perform tight-binding calculations of Raman spectra in double-layer graphene for a series of commensurate structures with varying rotational angles. We compute the Raman spectrum by standard perturbation expansion [17] in the electron-photon and electron-phonon interaction. To construct tight-binding models for a rotated double-layer graphene at arbitrary $\theta$, we use Slater-Koster parameters fitted to density functional theory (DFT) calculation of rotated graphene [3]. We also rescale all tight-binding hopping parameters by 18% in order to account for GW computed correlation effects [32]. For electron linewidth we use constant value of 190 meV (half width at half maximum) in order to reproduce correctly the amount of G-peak enhancement at the critical angle for 1.96 eV laser [33]. We expand the electron-photon and electron-phonon interaction to nearest neighbor carbon atoms. Unlike the electronic structure, which we compute for each rotational angle, we assume that the phonon band structure is unchanged going from single to double-layer graphene [28].

The computed Raman G and 2D peak features are consistent with the observed experimental trends. As shown in Fig. 2b and 2c, the calculated G peak intensities with two different laser energies (1.96 eV and 2.41 eV) show good agreements between experimentally observed and calculated critical angles. Our calculations also show that the main contribution to the G peak enhancement arises from electron-hole excitations at regions of strong inter-layer coupling (i.e., near the Van Hove singularities) (Fig. 4d) as one would expect from our simple model based on critical mismatch angle.

Computed values (red circles in Fig. 3) of 2D Raman spectra also agree well with experimentally obtained results, which confirms that the systematic trend in the behaviour of the 2D Raman peak originates from the changes in the electronic structure of rotated double-layer graphene. In order to further elucidate the origin of the 2D Raman features, we perform Raman spectrum calculations in which the effects of interlayer coupling are included only in the electron wavefunctions or only in the electron band energies. We find that double-layer interaction only in the electron band energies is enough to explain the dependence of the 2D Raman peak position on the rotational angle. Since the main change in electron band energies is associated with the opening of the gap, leading to before mentioned Van Hove singularities, we conclude that dependence of 2D peak position on the misorientation angle is a measure of location of the Van Hove singularities. The origin of the increased 2D FWHM and decreased intensity at the low misorientation angle regime is much more complex since the effect of interaction between layers on electron wavefunctions becomes important as well as the occurrence of constructive and destructive interference between various Raman scattering pathways.   We consider these effects in more detail in [36].   We have also performed a continuum model study of the Raman spectroscopy of rotated bilayers, using the reduced Hilbert space of Refs. [1,5].   The FWHM, peak position and intensity of the 2D Raman peak in the continuum model study show similar trends as those in the tight-binding model, thus re-enforcing our physical picture obtained here.

Finally, we note that the 2D peak features are also laser-energy dependent.   With higher excitation laser energies, we observe all the 2D peak features in peak width, blueshift, and intensity, are relatively shifted to higher misorientation angles, compared to data obtained from 1.96 eV laser energy [28].   Our calculation with 2.41 eV also correctly predicts the features in 2D peak data, confirming the validity of our analysis [28].

The standard Raman signature of single-layer and AB-stacked bilayer graphene [19,20] has been widely used to characterize chemically synthesized graphene samples where rotational stacking is common [26,37].   The present study shows that the Raman method for assignment of layer number should be applied with caution for rotationally stacked graphene samples.   Moreover, our study provides a convenient way to deduce the rotational angle in double-layer graphene.   From the combined information of the 2D peak width, 2D peak location, and the 2D/G intensity ratio, one can deduce the general rotational angles in double layer graphene samples.   Furthermore, using different laser excitation energies allows tuning of the critical transition angles [28].

Acknowledgements – We thank Feng Wang and Gregory Samsonidze for helpful discussions and Francesco Mauri for sharing data on calculated monolayer graphene phonon bandstructure.   This research was supported in part by the Director, Office of Energy Research, Materials Sciences and Engineering Division, of the US Department of Energy under contract No. DE-AC02-05CH11231 which provided for TEM characterizations and tight-binding calculations; by the National Science Foundation within the Center of Integrated Nanomechanical Systems, under Grant No. EEC-0832819, which provided for CVD graphene synthesis; by the National Science Foundation under grant No. 0906539 which provided for suspended double-layer graphene sample preparation and TEM analysis; by the National Science Foundation under grant No. DMR10-1006184 which provided for continuum model calculations; and by the Office of Naval Research (MURI) which provided for Raman spectroscopy and data analysis.    Portions of the present study were performed at the National Center for Electron Microscopy, Lawrence Berkeley National Laboratory, which is supported by the U.S. Department of Energy under Contract # DE-AC02-05CH11231.

**Figures**

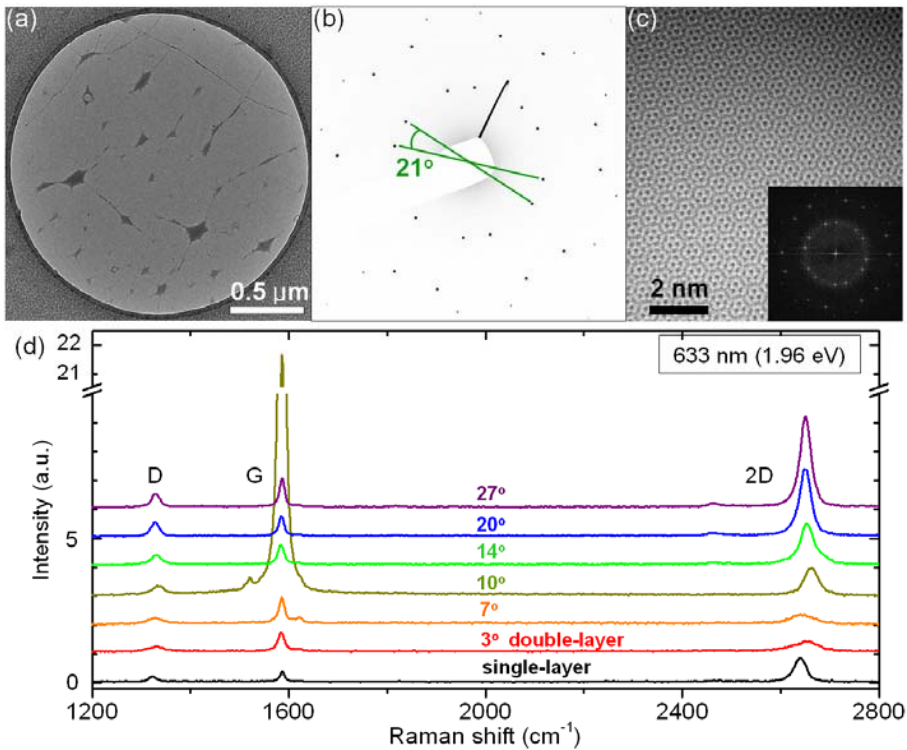

**FIG. 1. (Color online) Suspended rotated double-layer graphene and Raman spectrum.** (a) TEM image of misoriented double layer graphene. The graphene sample is suspended in a hole of 2 um diameter. (b) Diffraction pattern of the graphene sample shown in (a). The two sets of hexagonal patterns are relatively rotated by 21 degrees. An electron beam size of ~ 1 μm is used for diffraction acquisition. (c) Atomic resolution TEM image of double-layer graphene with rotational angle of 21 degree showing a moiré pattern. Inset shows a fast Fourier transform (FFT) of the image. (d) Raman spectra of misoriented double-layer and single layer graphene measured with 633 nm wavelength laser (1.96 eV). The spectra are shifted vertically for clarity.

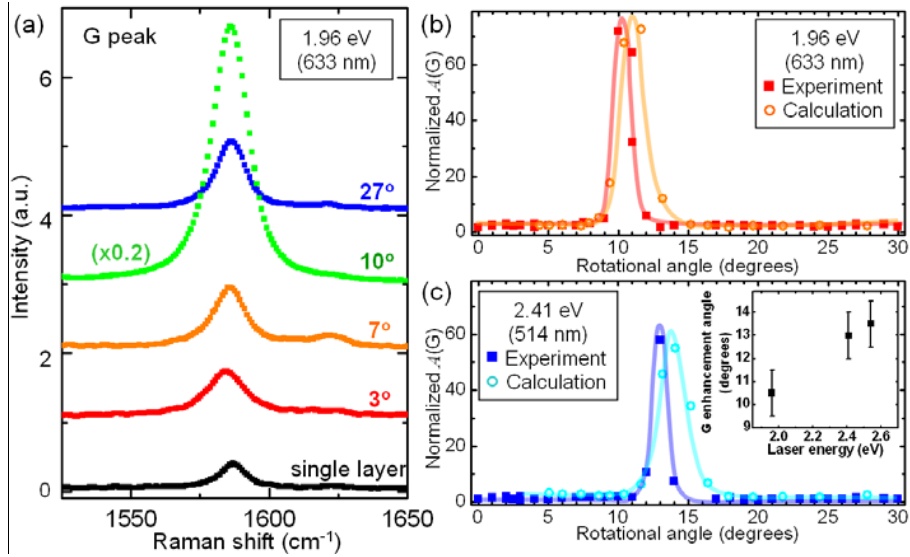

**FIG. 2. (Color online) Rotational angle dependence of G peak intensity.** (a) Graphene Raman G peak for rotated double-layer and single-layer graphene. (b) Dependence of G peak integral intensity (normalized to the single-layer value) on the rotation angle for 1.96 eV (633 nm) laser excitation. The filled and unfilled symbols show experimental and theoretical calculation values respectively. The lines are guides to the eye. (c) G peak integral intensity with 2.41 eV (514 nm) laser excitation. The inset shows measured dependence of critical angle on laser energy.

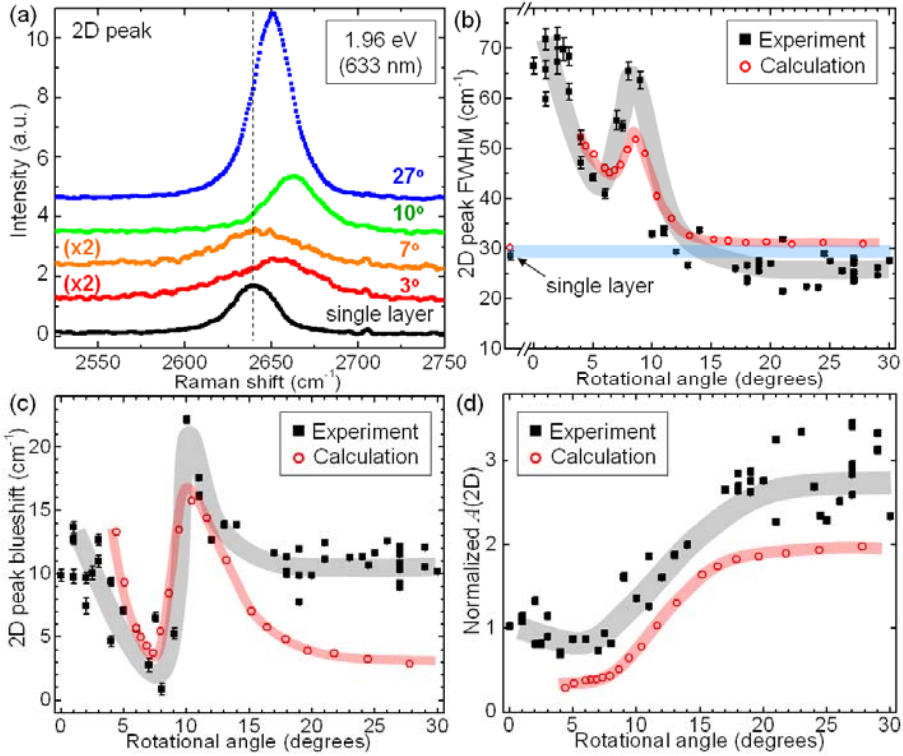

**FIG. 3. (Color online) Rotational angle dependence of Raman 2D peak.** (a) Graphene 2D peak for rotated double-layer and single-layer graphene. The vertical dashed line represents the center of single layer 2D peak. (b) Rotated double-layer graphene 2D peak FWHM. We have fitted the 2D peaks with a single Lorentzian peak for simplicity. The black squares and red circles are the experimental and theoretical calculation values. The blue horizontal area represents the experimental value from single layer graphene. The grey (experiment) and red (calculation) areas are guides to the eye. (c) Rotated double-layer graphene 2D peak blue-shift in respective to the value from single layer graphene. (d) Integral intensity of 2D peak. Experimental and calculation values were normalized to the single-layer value.

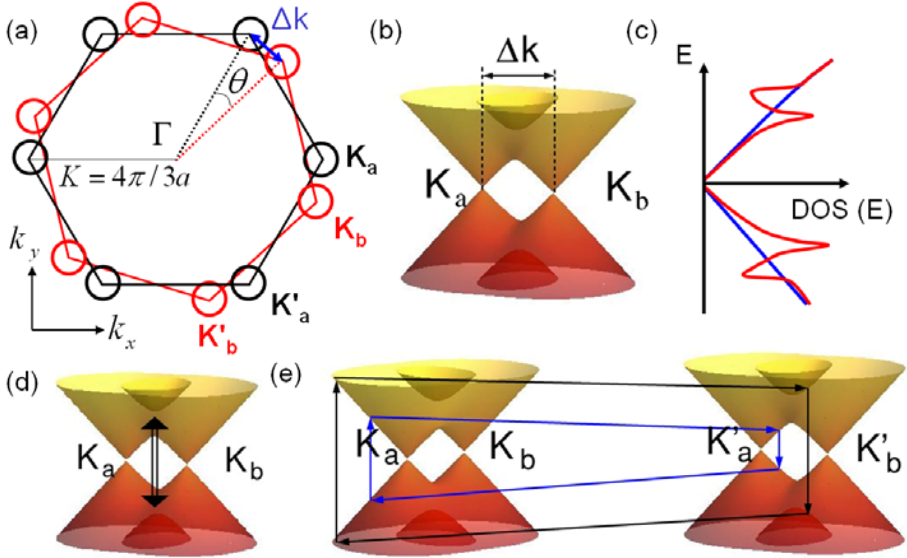

**FIG. 4. (Color online) Electronic band-structure and Raman scattering processes in rotated double-layer graphene.** (a) Brillouin zone (BZ) of rotated double-layer graphene misoriented by $\theta$. The circles represent the locations of Dirac cones from the first (black) and second (red) layer. The distance between two near-by Dirac cones is $\Delta k$. (b) Energy dispersion relation in the vicinities of two Dirac cone overlap. Van Hove singularities are induced from band-overlap between two Dirac cones. (c) Sketch of energy dependence of density of states (DOS) near the Fermi level of rotated double-layer graphene without (blue) and with interlayer interactions (red). DOS exhibit distortions from the interlayer interactions showing Van Hove singularities. (d) Some processes that contribute to the Raman G peak amplitude. (e) Intervalley 2D Raman scattering processes for rotated double-layer graphene in which the laser excitation energy is smaller (blue lines) or larger (black lines) than the energy difference between conduction and valence Van Hove singularities.

**Supplementary Material for**

# Raman spectroscopy study of rotated double-layer graphene:

# Misorientation-angle dependence of electronic structure

*Kwanpyo Kim[1,2,3], Sinisa Coh[1,3], Liang Z. Tan[1,3], William Regan[1,3], Jong Min Yuk[1,3,4], Eric Chatterjee[1], M. F. Crommie[1,2,3], Marvin L. Cohen[1,2,3], Steven G. Louie[1,3], and A. Zettl[1,2,3]*

[1]Department of Physics and [2]Center of Integrated Nanomechanical Systems, University of California at Berkeley, [3]Materials Sciences Division, Lawrence Berkeley National Laboratory, Berkeley, CA 94720,U.S.A. and [4]Department of Materials Science and Engineering, KAIST, Daejeon 305-701, Korea.

- Contents -

1. Sample preparation

2. TEM analysis of single layer graphene

3. TEM analysis of rotated double-layer graphene

4. Raman spectroscopy

5. Details of tight-binding calculation

Supplementary Figures S1-S6

References

**1. Sample preparation**

Rotated double-layer graphene samples are fabricated by stacking two layers of chemical vapor deposition (CVD)-grown graphene on Quantifoil holey carbon TEM grids.   After CVD graphene synthesis [S1], we perform a direct transfer method for transferring CVD graphene to TEM grids [S2]. Then we repeat the graphene transfer process to obtain the double-layer samples [S2,S3]. Supplementary Figure S1 show the optical and low-magnification TEM images of a sample. The TEM grid has an array of holes with 2 μm diameters.

**2. TEM analysis of single layer graphene**

Supplementary Figure S2 shows the TEM image and diffraction analysis of single-layer graphene. Our CVD-grown graphene is generally clean with some minor residues (Fig. S2a).   Diffraction pattern shows a hexagonal spot pattern (Fig. S2b) and inner (0-110) and outer (1-210) spot intensities show 1:1 ratio (Fig. S2c,d).   This confirms that the region is composed of single-layer graphene [S4,S5].   An electron beam size of ~ 1 μm is used for diffraction acquisition.

**3. TEM analysis of rotated double-layer graphene**

With TEM imaging and diffraction, we analyze the relative rotation angles between stacked graphene layers.   We check whether the graphene has tilt grain boundaries [S4] or local folds [S6], which prevents us from assigning well-defined rotation angles.   We also minimize the electron beam damage to graphene samples by performing only low magnification imaging and diffraction with low electron beam intensity.   Supplementary Figure S3 show the TEM image and diffraction analysis of rotated double-layer graphene.   Double-layer graphene has scattered dark regions, which are most likely carbon residues trapped between two layers (Fig. S3a).   These dark regions can be used to quickly distinguish double-layer from single-layer graphene regions in our samples.   If the double layer graphene has no tilt grain boundary or fold, diffraction patterns exhibit two sets of hexagonal patterns from two rotated graphene crystals as shown in Fig. S3b.   Analysis shows that the two rotated patterns have the similar intensities and maintain the single-layer signature, respectively.   This confirms that the region is composed of two stacked, rotated single-crystal graphene layers (Fig. S3c,d). Misorientation angles can be defined between zero and thirty degrees due to graphene hexagonal symmetry (Fig. S3e).   Regular TEM imaging and diffraction acquisition were performed in a JEOL 2010 microscope at 100 kV.   Atomic resolution TEM images are also acquired at 80 kV with a TEAM 0.5 microscope, at the National Center for Electron Microscopy, Lawrence Berkeley National Laboratory after Raman measurements.

## 4. Raman spectroscopy

All the Raman spectroscopy measurements were performed with a Renishaw inVia Raman microscope in ambient pressure at room temperature. 633 nm (1.96 eV), 514 nm (2.41 eV), and 488 nm (2.54 eV) laser wavelengths (energies) are used for measurements. A laser beam size of ~ 1 μm with ×100 objective lens is used. All the measurements were performed with low power (< 0.2 mW) to prevent a heating effect on the suspended graphene samples.

## 5. Details of tight-binding calculation

The Raman intensities are calculated from the tight-binding band structure and electronic wavefunctions by summing over terms in a perturbation expansion in the electron-photon and electron-phonon interactions. For the G peak we need to expand to third order, since it involves only one electron-phonon and two electron-photon interactions, while for 2D peak we need to expand to fourth order since it involves two phonon processes.

The primitive unit cell of a rotated double-layer graphene is much larger than that of a single layer graphene having only two carbon atoms (smallest unit cells are 7, 13, 19, and 31 times larger in size). For this reason, the BZ of a rotated double-layer graphene is reduced in size and any given k vector in the rotated double-layer BZ corresponds to a set of **k+G$_i$** vectors in the corresponding top and bottom BZ. Here **G$_i$** is a set of reciprocal vectors of the rotated double layer BZ chosen so that the **k+G$_i$** vectors for all **k** cover both the top and bottom monolayer reciprocal space exactly once.

As explained in the main manuscript, we assume that the phonon band structure of the rotated graphene layers is identical to that of monolayer graphene, and we simply fold it from single-layer BZ into a double-layer BZ. We use computed single-layer phonon bandstructure from Ref. [S7]. Furthermore, we approximate phonon eigenvectors at an arbitrary point near K (or K') with that at K (or K') point.

In order to further elucidate the origin of the 2D Raman features, we compare our Raman spectrum calculations with two different fictitious calculations. In the first, all electronic eigenenergies in the Raman intensity sum are replaced by the corresponding values in monolayer graphene, using the electronic wavefunctions of the rotated graphene layers. In the second, the values of the electronic wavefunctions in the Raman intensity sum are instead that of the graphene monolayer, while the eigenenergies are that of the rotated graphene layers.

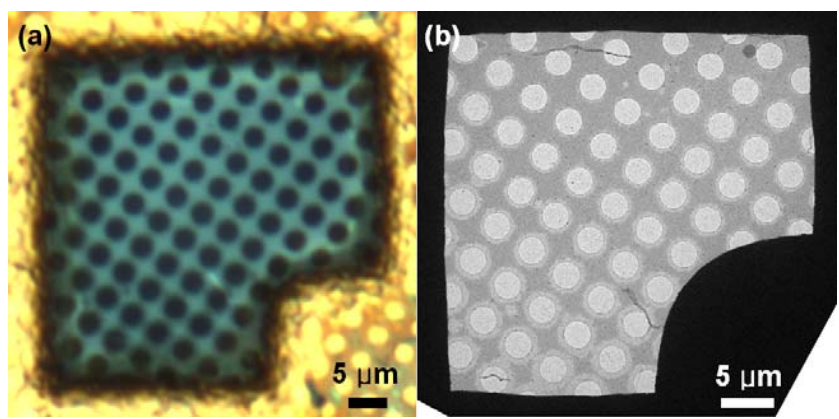

**Supplementary Figure S1. Double-layer graphene samples on a TEM grid.** (a) Optical image of a sample. (b) Low-magnification TEM image of the same regions of the sample.

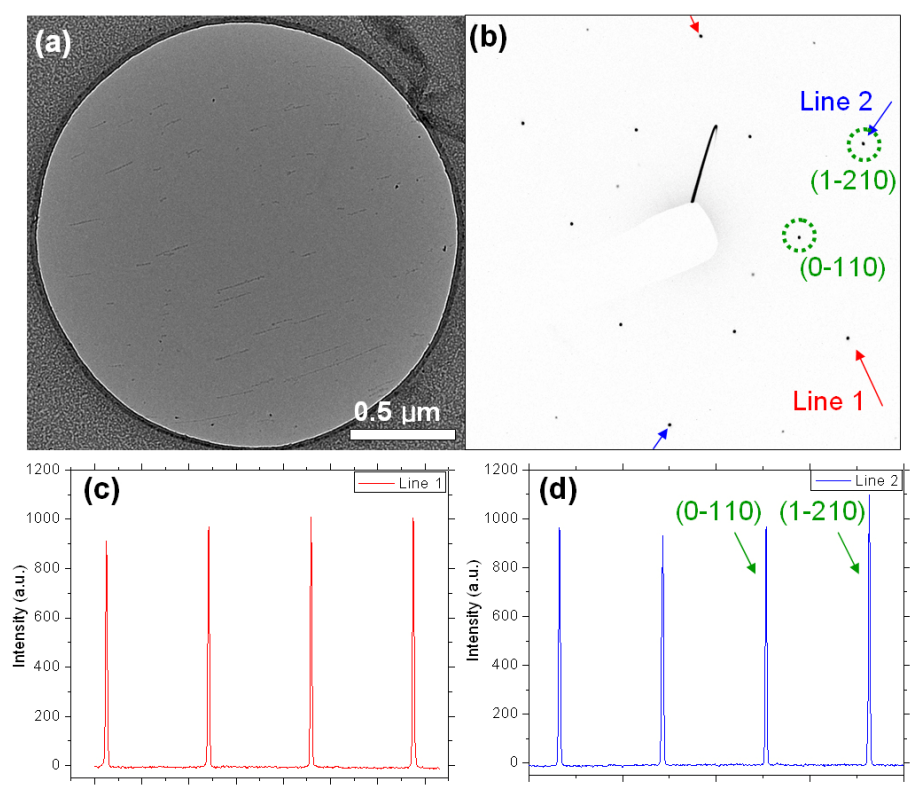

**Supplementary Figure S2. TEM imaging and diffraction of single layer graphene.** (a) TEM image of the suspended monolayer CVD graphene. The diameter of the hole is 2 μm. (b) Diffraction pattern of the center region in the hole. The diffraction shows the hexagonal patter. The arrows indicate the lines for intensity scan for Fig. c and d. The diffraction is acquired with about 1 μm e-beam size. (c) Line intensity of line 1. The intensities of inner and outer spot are comparable, showing that the graphene is monolayer. (d) Line intensity of line 2.

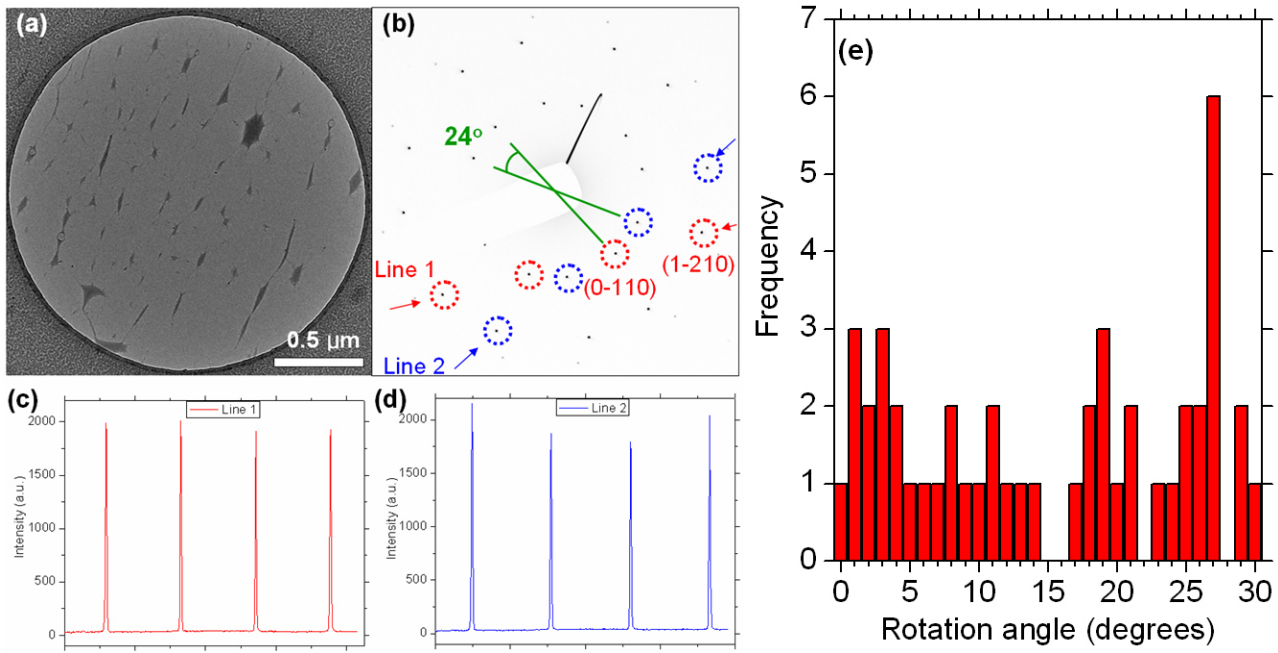

**Supplementary Figure S3. TEM imaging and diffraction of double layer graphene.** (a) TEM image of double layer graphene. (b) Diffraction patter of the graphene sample shown in Fig a. It shows 24 degree rotation. The arrows indicate the lines for intensity scan for Fig. c and d. The diffraction is acquired with about 1 μm e-beam size. (c) Line intensity of line 1. (d) Line intensity of line 2. (e) Histogram of rotation angles in the studied double-layer graphene samples.

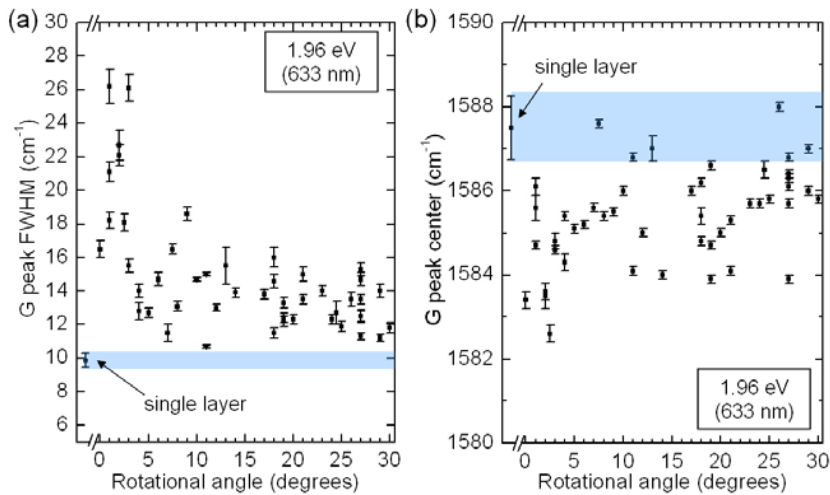

**Supplementary Figure 4. Rotational angle dependence of Graphene G peak FWHM and center location measured with 633 nm laser wavelength.** (a) Rotated double-layer graphene G peak full-width half-maximum (FWHM). The blue horizontal area represents the value from single layer graphene. (b) Rotated double-layer graphene G peak center location. The blue horizontal area represents the value from single layer graphene.

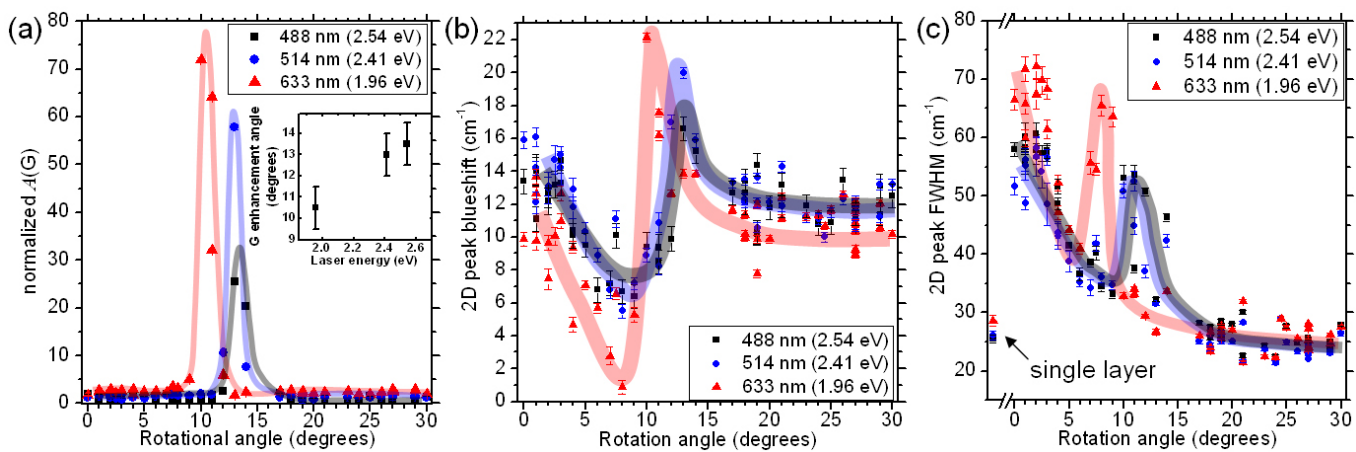

**Supplementary Figure S5. Experimental Raman data on rotated double-layer graphene with different laser wavelengths.** The general features in the data are shifted to higher rotational angles with higher laser energies. (a) G peak intensity depending on the rotation angle. Intensities were normalized to the single-layer value for each laser wavelength. The inset shows the laser energy dependence of rotational angles with G enhancement. (b) 2D peak blue-shift depending on the rotation angle compared to single-layer graphene values. (c) Graphene 2D peak FWHM depending on the rotation angle.

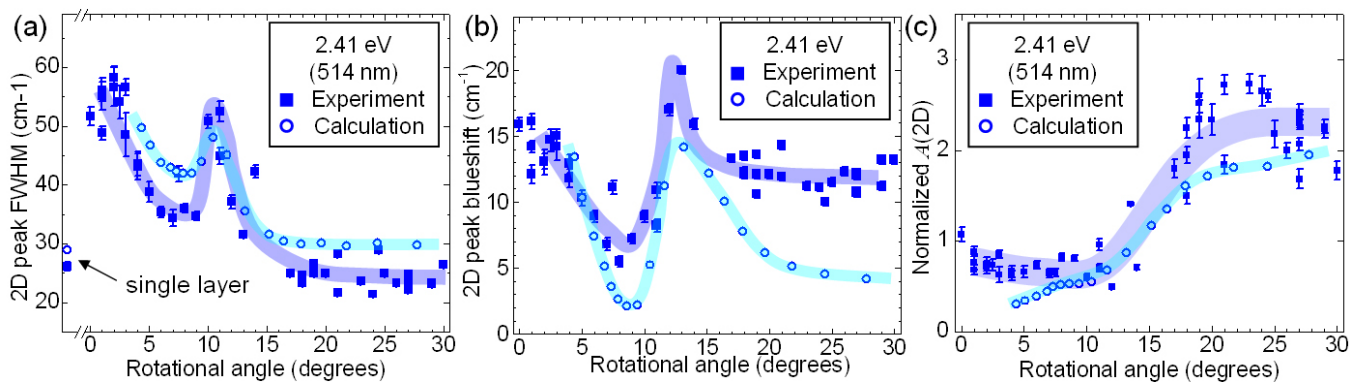

**Supplementary Figure S6. Rotational angle dependence of Raman 2D peak with 514 nm laser wavelength (2.41 eV).** (a) Rotated double-layer graphene 2D peak FWHM. The squares and circles are the experimental and theoretical calculation values. The blue (experiment) and bright blue (calculation) lines are guides to the eye. (b) Rotated double-layer graphene 2D peak blue-shift with respective to the value from single layer graphene. (c) Integral intensity of 2D peak. Experimental and calculation values are normalized to the single-layer value.